\documentclass[aps,prb,amssymb,twocolumn,showpacs]{revtex4}

\newcommand{\bk}{{\bf k}}
\newcommand{\bkp}{{{\bf k}^{\prime}}}
\newcommand{\bq}{{\bf q}}

\newcommand{\br}{{\bf r}}

\newcommand{\vf}{v_{\rm F}}

\newcommand{\be}{\begin{equation}}
\newcommand{\ee}{\end{equation}}
\newcommand{\bea}{\begin{eqnarray}}
\newcommand{\eea}{\end{eqnarray}}

\newcommand{\bwt}{\begin{widetext}}
\newcommand{\ewt}{\end{widetext}}

\usepackage{graphics,graphicx,amsmath}
\usepackage{amssymb}
\usepackage{tabularx,float}
\usepackage{epsfig}
\usepackage{bm,bbold}

\usepackage{color}
\usepackage{verbatim}

\begin{document}

\bibliographystyle{apsrev}

\date{\today}

\author{Rafael Rold\'{a}n and Luis Brey}
\affiliation{Instituto de Ciencia de Materiales de Madrid, CSIC, Cantoblanco E28049 Madrid, Spain}

\title{Dielectric screening and plasmons in AA-stacked bilayer graphene}

\begin{abstract}
The screening properties and collective excitations (plasmons) in AA-stacked bilayer graphene are studied within the random phase approximation (RPA). Whereas long lived plasmons in single layer graphene and in AB-stacked bilayer graphene can exist only in doped samples, we find that coherent plasmons can disperse in AA-stacked bilayer graphene {\it even in the absence of doping}. Moreover, we show that the characteristic low energy dispersion relation is unaffected by changes in the number of carriers, unless the chemical potential of the doped sample exceeds the inter-layer hopping energy. We further consider the effect of an external electric field applied perpendicular to the layers, and show how the dispersion of the modes can be tuned by the application of a gate voltage.  
\end{abstract}


\maketitle

\section{Introduction}

The unique optical and electronic properties of graphene have made this material an optimal candidate for plasmonics applications.\cite{GPN12} This possibility has brought a great interest on understanding the collective excitations in this material. Plasmons, which can be defined as collective density oscillations of an electron liquid, are present in many metals and semiconductors.\cite{AFS82,GV05} The characteristic linear dispersion relation of the quasiparticles in graphene, makes that plasmons in this material are manifestly different than in standard two-dimensional electron gases (2DEG) with a parabolic band dispersion.\cite{S86,WSSG06,HS07,P09,RGF10} In the absence of doping, no phase space in the excitation spectrum of single layer graphene (SLG) is available for the dispersion of plasmons. Something similar happens for bilayer graphene with the more stable Bernal or AB stacking (AB-BLG).\cite{SHS10,WC10,G11,YRK11,SG12} The recent progress on growing stable BLGs with AA stacking\cite{LI09,BSP11} has opened the possibility to study the properties of this class of graphene,\cite{A11,PF11,TN12,CR12,BF13} in particular the collective excitations. In this kind of stacking, one atom in the top layer belonging to the A(B) sublattice is directly adjacent to an A(B) atom of the bottom layer.\cite{A11,TN12} 

In undoped AA-stacked bilayer graphene, the perfect nesting of the electron and hole pockets leads, in presence of electron-electron interaction, to symmetry breaking ground states, such as antiferromagnetism or charge density wave.\cite{RN12,BF13} The transition temperature of these broken symmetry phases has been estimated to be of the order of a few degrees Kelvin,\cite{BF13} and at the present there is not experimental evidence for such gapped phases. In this work we consider temperatures higher than the transition temperatures  and assume that the ground state of the AA-stacked bilayer graphene is an uniform and paramagnetic free Dirac-like electron gas.

In this work we study the excitation spectrum and the collective modes of AA-BLG. The static screening properties are discussed and compared to those of a SLG. Analytical expressions for the low energy dispersion relations are given. We show that, contrary to SLG and AB-BLG, long lived plasmons exist in AA-BLG even in undoped samples. Furthermore, we find that due to the characteristic Drude weight in AA-BLG, the dispersion relation of the optical plasmon modes are independent of doping, unless the chemical potential of the sample exceeds the inter-layer hopping energy. By including an external electric field, applied perpendicular to the sample, we show that the velocity of the modes can be controlled by the strength of the gate voltage. 

The paper is organized as follows. In Sec. \ref{Sec:Hamiltonian} we analyze the band structure of AA-BLG. The polarization and dielectric functions within the RPA are studied in Sec. \ref{Sec:RPA}. The static screening properties are considered. Sec. \ref{Sec:CE} is devoted to the study of the collective excitations of the system. A discussion of our main results, stressing the main differences of the plasmon modes in AA-BLG with respect to other 2D systems, is done in Sec. \ref{Sec:Discussion}. Finally, our main conclusions are summarized in Sec. \ref{Sec:Conclusions}.

\section{Bilayer graphene with AA-stacking}\label{Sec:Hamiltonian}

\begin{figure}[t]
\begin{center}
\mbox{
\includegraphics[width=3.5cm]{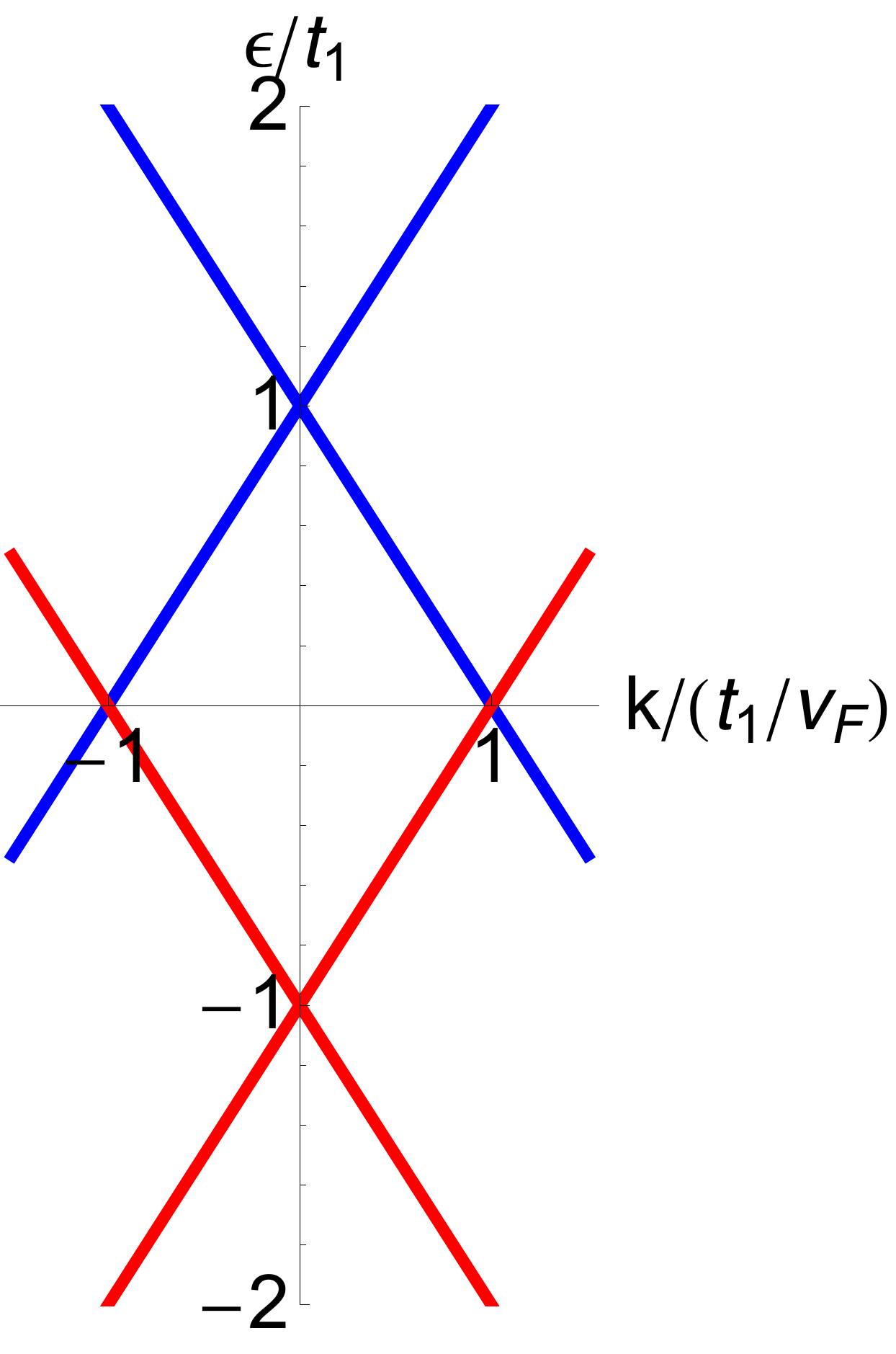}
\includegraphics[width=4.5cm]{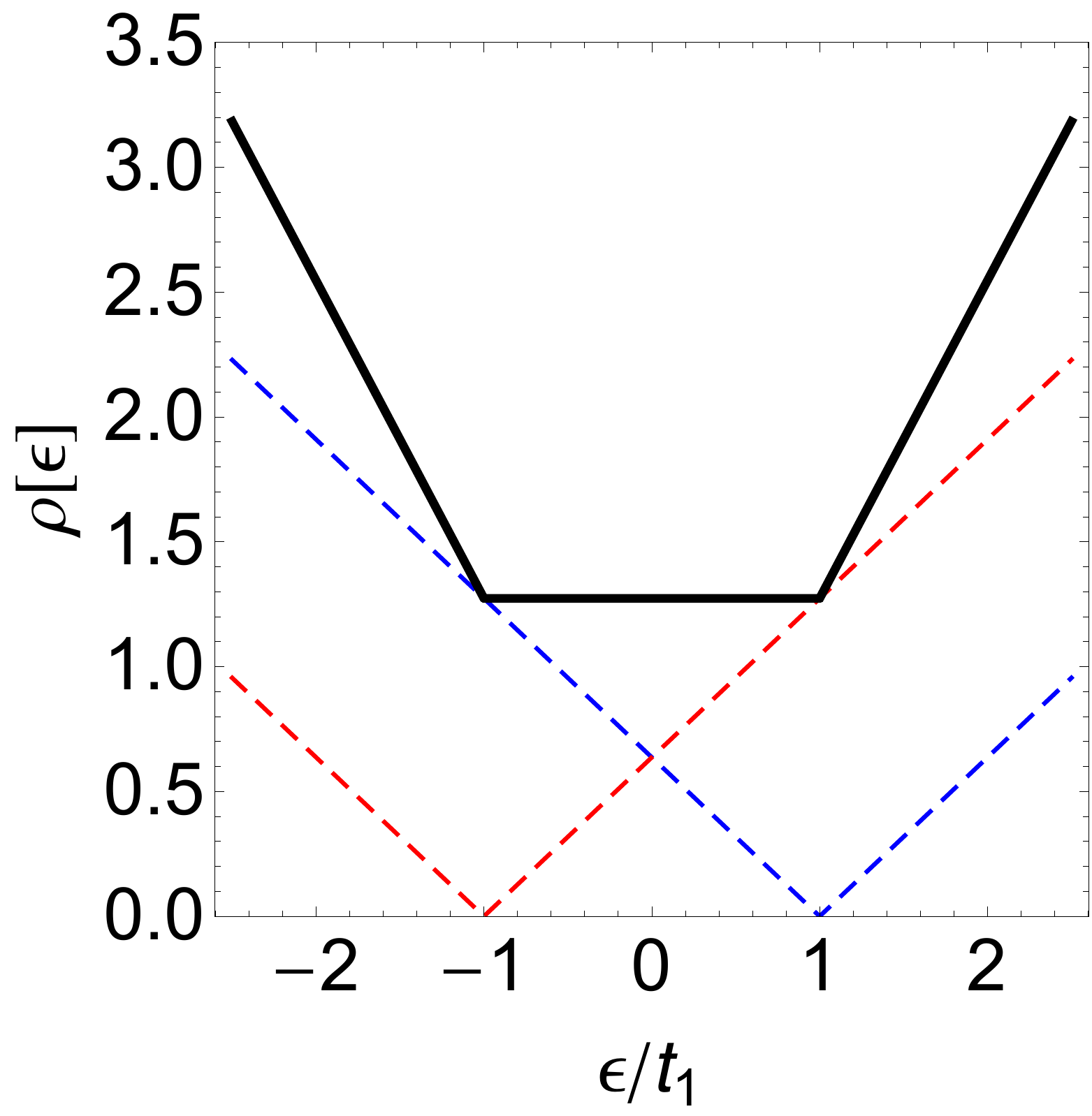}
} 
\end{center}
\caption{(Color online) Left: Dispersion relation of the AA-BLG calculated from Eq. (\ref{Eq:EnergyAA}). Blue lines corresponds to $s=+1$ and red lines to $s=-1$. Right: Density of states, as given by Eq. (\ref{Eq:DOS}). The contribution of each cone is shown by dashed lines, and the total DOS is given by the full black line.}
\label{Fig:EnergyAA}
\end{figure}

The unit cell of an AA-BLG consists of 4 inequivalent Carbon atoms, two for each layer. The two-dimensional Brilloin zone is an hexagon and, as in SLG, the low energy physics occurs near the two Dirac points $K$ and $K'$.\cite{CG09,K12} We describe the electronic properties of an AA-BLG by a single valley model for electrons in each layer, with an inter-layer hopping $t_1\sim 0.2$~eV and separation between the layers, $d\sim 3.6$~\AA.\cite{XLD10,LP11} The low energy Hamiltonian around the $K$-point can be expressed as
\be
 H(\bk)
=
\left( 
\begin{array}{cc} 
H_{{\rm SL}}(\bk) & H_{\perp} \\
 H_{\perp}  & H_{{\rm SL}}(\bk) 
\end{array} 
\right)
\label{Eq:HAA}
\ee
where 
\be
{\cal H}_{{\rm SL}}(\bk)
=
\vf\left( 
\begin{array}{cc} 
0 & k_{-}  \\
k_{+} & 0 
\end{array} 
\right)
\ee
is the usual single-layer graphene Hamiltonian, where $v_F$ is the Fermi velocity, $k_{\pm}=k_x\pm ik_y$, and 
\be\label{Eq:Hperp}
{\cal H}_{\perp}
=
\left( 
\begin{array}{cc} 
 -t_1 & 0 \\
 0 & -t_1 
\end{array} 
\right)
\ee
corresponds to the inter-layer hopping, which for this kind of stacking, accounts for tunneling processes between Carbon atoms with the same sublattice indices. By applying the transformation $U^{-1}HU$, where 
\be
U
=
\left( 
\begin{array}{cccc} 
1 & 0 & -1 & 0 \\
 0 & 1 & 0 & -1 \\
 1 & 0 & 1 & 0 \\
 0 & 1 & 0 & 1
\end{array} 
\right)
\label{Eq:U}
\ee
we can rewrite the Hamiltonian (\ref{Eq:HAA}) in the block diagonal form
\be
{\cal H}(\bk)
=
\left( 
\begin{array}{cccc} 
-t_1 & v_Fk_{-} & 0 & 0 \\
 \vf k_{+} & -t_1 & 0 & 0 \\
 0 & 0 & t_1 & \vf k_{-} \\
 0 & 0 & \vf k_{+} & t_1
\end{array} 
\right).
\label{Eq:HAA2}
\ee

Eq. (\ref{Eq:HAA2}) accounts for two uncoupled set of bands, which admits the eigen-energies
\be\label{Eq:EnergyAA}
\varepsilon_{\lambda,s}(\bk)=st_1+\lambda\vf k
\ee
where $s=\pm 1$ labels the decoupled {\it bonding/antibonding} blocks of Eq. (\ref{Eq:HAA2}), $\lambda=\pm 1$ is the band index, and $k=(k_x^2+k_y^2)^{1/2}$. The band structure obtained from (\ref{Eq:EnergyAA}) is shown in Fig. \ref{Fig:EnergyAA}. The eigenfunctions take the form
\be
\Psi_{s,\lambda}(\bk)
=
\left(
\begin{array}{c}
 \psi_{\lambda}(\bk)\\
0
\end{array}
\right)
\hspace{0.5cm}
{\rm and}
\hspace{0.5cm}
\left(
\begin{array}{c}
0\\
 \psi_{\lambda}(\bk)
\end{array}
\right)
\ee
for $s=\pm 1$, respectively, where
\be\label{Eq:psi_lambda}
\psi_{\lambda}(\bk)
=
\frac{1}{\sqrt{2}}
\left( 
\begin{array}{c} 
\lambda e^{i\phi_{\bk}} \\ 
1 
\end{array} 
\right)
\ee
and $\tan \phi_{\bk}=k_y/k_x$. The low energy density of states of AA-BLG is\cite{A11,TN12}
\be\label{Eq:DOS}
\rho(\varepsilon)=\frac{g_{\sigma}g_vt_1}{2\pi \vf^2}\left( \left|\frac{\varepsilon}{t_1}-1 \right|+\left|\frac{\varepsilon}{t_1}+1 \right|\right)
\ee
where $g_{\sigma}g_v$ is the spin and valley degeneracy. The DOS contains contributions from the bonding and antibonding blocks, as it is shown in Fig. \ref{Fig:EnergyAA}. Notice that $\rho(\varepsilon)={\rm cte.}$ at low energies, as in a standard 2DEG, whereas for $|\varepsilon|>t_1$ it grows linearly with energy, as in SLG.

\section{Polarization function and the Random-Phase Approximation}\label{Sec:RPA}
Since we are interested on studying collective modes, we need to calculate the dielectric function ${\bm\epsilon}(\bq,\omega)$ of the system. Within the RPA, ${\bm\epsilon}(\bq,\omega)$ is expressed as:
\be\label{Eq:EpsilonRPA}
{\bm\epsilon}(\bq,\omega)=   {\mathbb 1}-{\bm V}(q)\cdot{\bm {\hat{\bm{\Pi}}}}^0(\bq,\omega),
\ee
where ${\bm\epsilon}(\bq,\omega)$, ${\bm V}(\bq)$ and ${\bm {\hat{\bm{\Pi}}}}^0(\bq,\omega)$ are $2\times 2$ matrices. ${\bm V}(q)$ is the Coulomb interaction matrix, and ${\bm {\hat{\bm{\Pi}}}}^0(\bq,\omega)$ is the polarization matrix, whose elements are the density-density linear-response functions. In our problem, ${\bm V}(q)$ accounts for the intra-layer repulsion between electrons within the same graphene sheet, through the 2D Fourier transform of the long range Coulomb interaction
\be
V_{\rm intra}(q)=\frac{2\pi e^2}{\kappa q}
\ee
and for the inter-layer interaction between electrons in different layers
\be
V_{\rm inter}(q)=V_{\rm intra}(q)e^{-qd}
\ee
where $d$ is the separation between the layers and $\kappa$ is the dielectric constant of the embedding medium. In order to work in the same basis as in (\ref{Eq:HAA2}), we transform the Coulomb interaction matrix from the layer$_1$/layer$_2$ basis
\be
{\tilde{\bm V}}(q)
=
\left(
\begin{array}{cc} 
V_{\rm intra}(q) & V_{\rm inter}(q) \\
 V_{\rm inter}(q) & V_{\rm intra}(q) 
\end{array} 
\right)
\ee
to the {\it bonding/antibonding} basis through the transformation
$
{\bm V}(q)={\cal U}^{-1}{\tilde{\bm V}}(q){\cal U}
$
where
$
{\cal U}
=
\left(
\begin{array}{cc} 
1& -1 \\
 1 & 1
\end{array} 
\right).
$
This leads to
\be\label{Eq:V(q)Matrix}
{\bm V}(q)
=
\left(
\begin{array}{cc} 
V_{+}(q)  & 0\\
0&  V_{-}(q) 
\end{array} 
\right)
\ee
where
\be
V_{\pm}(\bq)=V_{\rm intra}(q) \pm V_{\rm inter}(q).
\ee

On the other hand, ${\bm {\hat{\bm{\Pi}}}}^0(\bq,\omega)$ contains elements of the form
\bea
\Pi^0_{s, s'; \lambda,  \lambda'  }(\bq,\omega)\hspace{6cm}&&\nonumber\\
=-\frac{g_{\sigma}g_v}{L^2}\sum_{\bk}\frac{f_{s,\lambda}(\bk)-f_{s',\lambda'}(\bk+\bq)}{\omega+\varepsilon_{s,\lambda}(\bk)-\varepsilon_{s',\lambda'}(\bk+\bq)+i\delta}&&\nonumber\\
\times{\cal F}_{ss';\lambda\lambda'}(\bk,\bk+\bq)\hspace{3.5cm}&&
\eea
where $L^2$ is the sample area and $f_{s,\lambda}(\bk)=[\exp\{\beta(\varepsilon_{s,\lambda}(\bk)-\mu)\}+1]^{-1}$ is the Fermi-Dirac distribution function, which for $\mu=0$ and $T=0$ reads $\Theta[\varepsilon_{s,\lambda}(\bk)]$, and $\delta=0^+$. It is important to notice that the overlap of the electron and hole wave-functions ${\cal F}_{ss';\lambda\lambda'}(\bk,\bkp)=|\langle\Psi_{s,\lambda}(\bk)|e^{i(\bk-\bkp)\cdot \br}|\Psi_{s',\lambda'}(\bkp)\rangle|^2$ is
\be\label{Eq:F}
{\cal F}_{ss';\lambda\lambda'}(\bk,\bkp)
 = \left\{ \begin{array}{ccc}
\frac{1+\lambda\lambda'\cos(\phi_{\bk}-\phi_{\bkp})}{2}& \mbox{for} & s=s' \\ 
  0  & \mbox{for} & s\ne s' 
\end{array}\right.
 \ee
 Because from Eq. (\ref{Eq:F}), ${\cal F}_{ss';\lambda\lambda'}(\bk,\bkp)=0$ for $s\ne s'$, only electron-hole transitions between bands with the same $s$ contribute to the polarizability of the AA-BLG. This implies that only transitions between symmetric bands (blue lines in Fig. \ref{Fig:EnergyAA}) or between antisymmetric bands (red lines) are allowed. Then, each cone contributes independently and the polarization can be expressed as
 \be\label{Eq:Pi0Matrix}
{\bm \hat{{\bm \Pi}}}^0(\bq,\omega)
=
\Pi^0(\bq,\omega)\left( 
\begin{array}{cc} 
1 & 0 \\
 0 & 1 
\end{array} 
\right)
\ee
where

\bea\label{Eq:Pi0}
\Pi^0(\bq,\omega)=-\frac{g_{\sigma}g_v}{L^2}\sum_{\lambda\lambda'}\sum_{\bk}\frac{f_{\lambda}(\bk)-f_{\lambda'}(\bk+\bq)}{\omega+\varepsilon_{\lambda}(\bk)-\varepsilon_{\lambda'}(\bk+\bq)+i\delta}&&\nonumber\\
\times{\cal F}_{\lambda\lambda'}(\bk,\bk+\bq).\hspace{4cm}&&
\eea

\begin{figure}[t]
\begin{center}
\mbox{
\includegraphics[width=7.5cm]{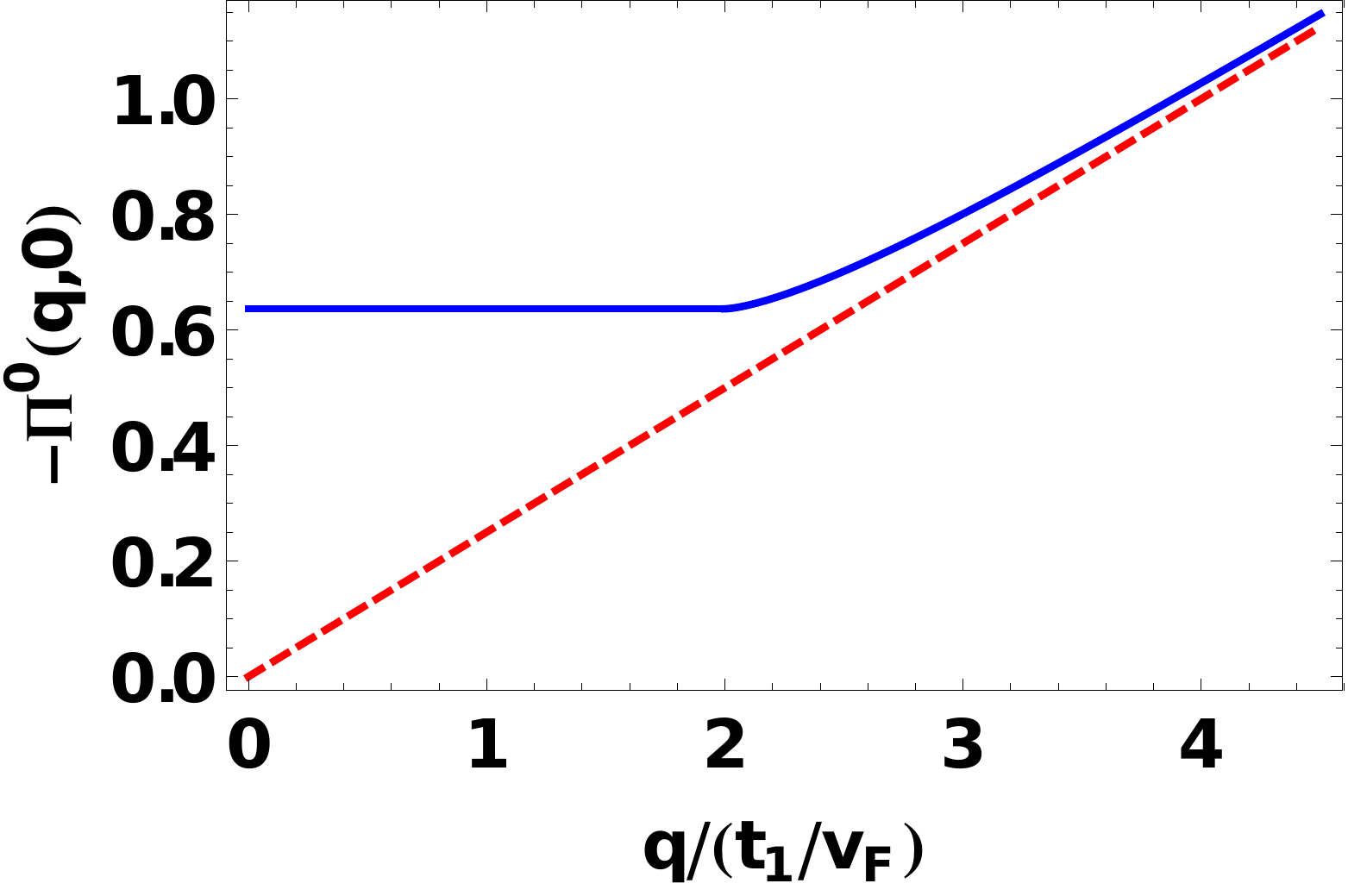}
} 
\end{center}
\caption{(Color online) Static polarization function $-\Pi^0(q,\omega=0)$ for AA-BLG (solid blue line). The dashed red line corresponds to the polarization in the limit of decoupled layers $t_1\rightarrow 0$, which coincides with the static polarization of undoped SLG.}
\label{Fig:Pi0Static}
\end{figure}

The remaining part of the calculation of the polarization Eq. (\ref{Eq:Pi0}) reduces to the polarization of a doped SLG with a finite chemical potential $\mu$,\cite{S86,WSSG06,HS07} by just substituting $\mu \longleftrightarrow t_1$.

As a first result, we discuss the static screening of AA-BLG. In the $\omega\rightarrow 0$ limit, the static polarization function $\Pi^0(q,\omega=0)$ is entirely real and behaves as shown in Fig. \ref{Fig:Pi0Static}. One notices that the structure of the static polarization function is similar to that of a doped SLG,\cite{HS07} containing two different contributions: a constant metallic-like and a linear in $q$ insulating-like screening regions. However, contrary to SLG, for which the metallic like screening contribution appears only in the doped regime and which is controlled by the chemical potential, AA-BLG shows a metallic like screening even in the absence of doping, due to the {\it always} finite density of states in the spectrum (see Fig. \ref{Fig:EnergyAA}). The polarization function for this region of wave-vectors is
\be
\Pi^0(q)=-\rho(0)=-\frac{g_{\sigma}g_vt_1}{\pi v_F^2}\hspace{0.5cm} {\rm for}\hspace{0.5cm} q\le 2t_1/v_F
\ee
which is a constant that depends only on the inter-layer hopping $t_1$ and on the Fermi velocity $v_F$.

At $q= 2t_1/v_F$ there is a crossover from metallic to insulating screening, the latter being associated to inter-band transitions which lead to a linear behavior of the polarization function, as shown in Fig. \ref{Fig:Pi0Static}. In the limit of two decoupled layers ($t_1\rightarrow 0$), one recovers the typical polarization function of an undoped SLG, as given by the red dashed line in Fig. \ref{Fig:Pi0Static}. The static screening in the long wavelength limit is simply given by $\varepsilon(q)\approx 1+q_{TF}/q$, where the Thomas-Fermi wave vector in this case is $q_{TF}=2\pi e^2\rho(0)/\kappa=2e^2g_{\sigma}g_vt_1/(\kappa v_F^2)$. Notice that $q_{TF}$ in AA-BLG, as in an standard 2DEG (and contrary to SLG), is independent of the carrier concentration. However, the static screening in a 2DEG falls off rapidly at large wave-vectors, whereas for both, SLG and AA-BLG,  it grows linearly due to the contribution from inter-band excitations.  Since the DOS (\ref{Eq:DOS}) is constant at low energies, the peculiar static screening discussed above will remain for doped AA-BLG, provided that $|\mu|\le t_1$.

\begin{figure}[t]
\begin{center}
\mbox{
\includegraphics[width=7.3cm]{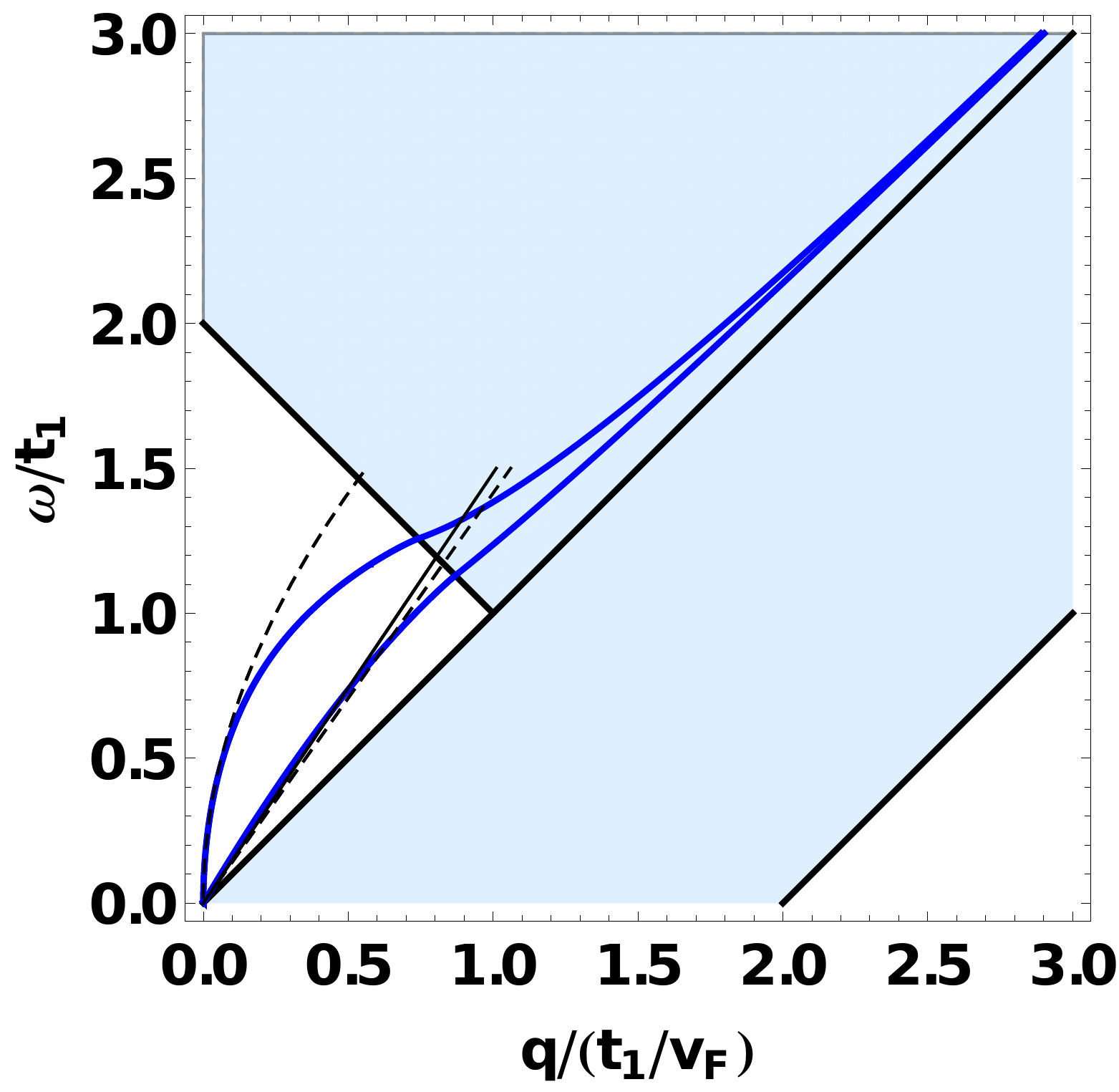}
} 
\end{center}
\caption{(Color online) The dispersion relation of the plasmon modes calculated from Eq. (\ref{Eq:Plasmons}) is shown by the solid blue lines. The approximated dispersion given by Eq. (\ref{Eq:PlasmonOpt}) and (\ref{Eq:PlasmonAc}) are shown by the dashed black lines. The dispersion relation of the acoustic mode given by (\ref{Eq:PlasmonAcFull}) is shown by the full black line. The thick black lines delimitate the intra-band and inter-band electron-hole continuum, which correspond the the colored regions of the plot.}
\label{Fig:Plasmons}
\end{figure}
 For finite frequencies, it is possible to obtain a closed analytical expression for the polarization function valid in the region of the spectrum where plasmons are undamped, this is, for $\omega>v_Fq$ and ${\rm Im}\Pi^0(\bq,\omega)=0$, which can be written as
\bea\label{Eq:Pi0Analytical}
\Pi^0(\bq,\omega)&=&-\frac{g_{\sigma}g_vt_1}{2\pi v_F^2}+\frac{g_{\sigma}g_v}{16\pi}\frac{q^2}{\sqrt{\omega^2-v_F^2q^2}}\nonumber\\
&\times&\left[G\left(\frac{2t_1+\omega}{v_Fq}\right)-G\left(\frac{2t_1-\omega}{v_Fq}\right) \right],
\eea
where $G(x)=x\sqrt{x^2-1}-\cosh^{-1}(x)$ for $x>1$. The electron-hole continuum of AA-BLG, defined as the region of the $\omega-q$ plane where electron-hole pairs can be excited,\cite{RGF10} can be easily calculated from ${\rm Im}~\Pi^0(q,\omega)$. Due to the peculiar band structure of this system, with two independent set of bands, the electron-hole continuum of AA-BLG has same appearance as the one of a doped SLG, but again interchanging the chemical potential $\mu$ by the inter-layer hopping $t_1$. Such spectrum, which contains two different regions, associated to intra- and inter-band excitations respectively, is shown by the colored region in Fig. \ref{Fig:Plasmons}. As a consequence, optical ($q=0$) transitions in undoped AA-BLG are Pauli blocked, impeding the absorption of photons with energy less than $2t_1$.

\section{Collective excitations}\label{Sec:CE}

In this section we study the dispersion relation of the plasmons in AA-BLG, which is obtained from the zeros of the equation
\be\label{Eq:Plasmons}
\det {\bm\epsilon}(\bq,\omega)=0.
\ee
From (\ref{Eq:V(q)Matrix}) and (\ref{Eq:Pi0Matrix}), it is easy to show that the condition (\ref{Eq:Plasmons}) leads to the following equations for collective modes:
\bea
\epsilon_+^{RPA}(\bq,\omega)&=&1-V_+(q)\Pi^0(\bq,\omega)=0\label{Eq:Plasmon+},\\
\epsilon_-^{RPA}(\bq,\omega)&=&1-V_-(q)\Pi^0(\bq,\omega)=0\label{Eq:Plasmon-}.
\eea

\subsection{Plasmons in neutral AA-BLG}\label{Sec:Plasmons}

The numerical solution for the collective modes obtained from Eq. (\ref{Eq:Plasmon+}) and (\ref{Eq:Plasmon-}) for the undoped system are shown by the blue lines in Fig. \ref{Fig:Plasmons}. As it is usual in bilayer systems, we obtain two modes: one {\it optical} mode, with a $\omega_+(q\rightarrow 0)\sim \sqrt{q}$ dispersion, and one {\it acoustic} mode with a linear dispersion relation $\omega_-(q\rightarrow 0)\sim q$. Whereas the optical mode $\omega_+(q)$ corresponds to a collective excitation in which the densities in the two layers fluctuate in phase, the acoustic plasmon $\omega_-(q)$ accounts for an out-of-phase oscillation of the carriers in the two layers. We can easily obtain an analytic expression for the low energy dispersion relation of the modes, by using the long-wavelength limit of the polarization function (\ref{Eq:Pi0Analytical})
\be\label{Eq:PiSim}
\Pi^0_{q\rightarrow 0}(\bq,\omega)=g_{\sigma}g_v\frac{t_1}{4\pi}\frac{q^2}{\omega^2},
\ee
from which we obtain the two plasmon branches $\omega_{\pm}(q)=\sqrt{\frac{(1\pm e^{-qd})e^2g_{\sigma}g_vt_1}{2\kappa}q}$, which have a low energy behavior
\be\label{Eq:PlasmonOpt}
\omega_{+}(q)\approx \sqrt{\frac{e^2g_{\sigma}g_vt_1}{\kappa}q}
\ee
and
\be\label{Eq:PlasmonAc}
\omega_{-}(q)\approx \sqrt{\frac{e^2g_{\sigma}g_vdt_1}{2\kappa}}\,\,q.
\ee
The approximations (\ref{Eq:PlasmonOpt}) and (\ref{Eq:PlasmonAc}) are shown by the dashed black lines in Fig. \ref{Fig:Plasmons}. Contrary to SLG or AB-stacked BLG, it is interesting to note the existence of long-lived plasmon modes in AA-BLG already in the undoped regime. In Sec. \ref{Sec:Doped} we will see that the dispersion relation of the optical plasmon (\ref{Eq:PlasmonOpt}), in the moderate doped regime, will be unaffected by induced charge carriers in the system. 

It is important to notice that the long wavelength limit of $\omega_{-}(q)$, as given by Eq. (\ref{Eq:PlasmonAc}), is only valid for $t_1d/(v_F\kappa) \gg 1$. As a consequence, the {\it exact} numerical solution for the dispersion relation of the acoustic mode is not properly fitted by Eq. (\ref{Eq:PlasmonAc}), as it can be seen by the dashed black line in Fig. \ref{Fig:Plasmons}, which lies below the RPA numerical solution, underestimating the velocity of the mode $v_s$. For the general case, as it is discussed in Refs. \onlinecite{SG88} and \onlinecite{PM12}, the dispersion relation of the linear acoustic mode must be calculated in terms of a Laurent-Taylor expansion including the full polarization function (\ref{Eq:Pi0Analytical}). The square root singularity of $\Pi^0(\bq,\omega)$ at $\omega=v_Fq$ ensures that $v_s>v_F$.\cite{SG12b} Following the scheme developed by Profumo et al. in Ref. \onlinecite{PM12}, we obtain the next expression for the acoustic plasmon dispersion
\be\label{Eq:PlasmonAcFull}
\omega_{-}(q)=\frac{1+\frac{g_{\sigma}g_v\sqrt{2}e^2dt_1}{2\kappa v_F^2}}{\left(1+\frac{g_{\sigma}g_v\sqrt{2}e^2dt_1}{\kappa v_F^2}\right)^{1/2}}\, v_Fq.
\ee
The approximated dispersion relation given by Eq. (\ref{Eq:PlasmonAcFull}) is shown by the full black line in Fig. \ref{Fig:Plasmons}, which clearly matches the RPA numerical solution (thick blue line) at small values of $q$. 

\subsection{Effect of a perpendicular electric field}

\begin{figure}[b]
\begin{center}
\mbox{
\includegraphics[width=4.2cm]{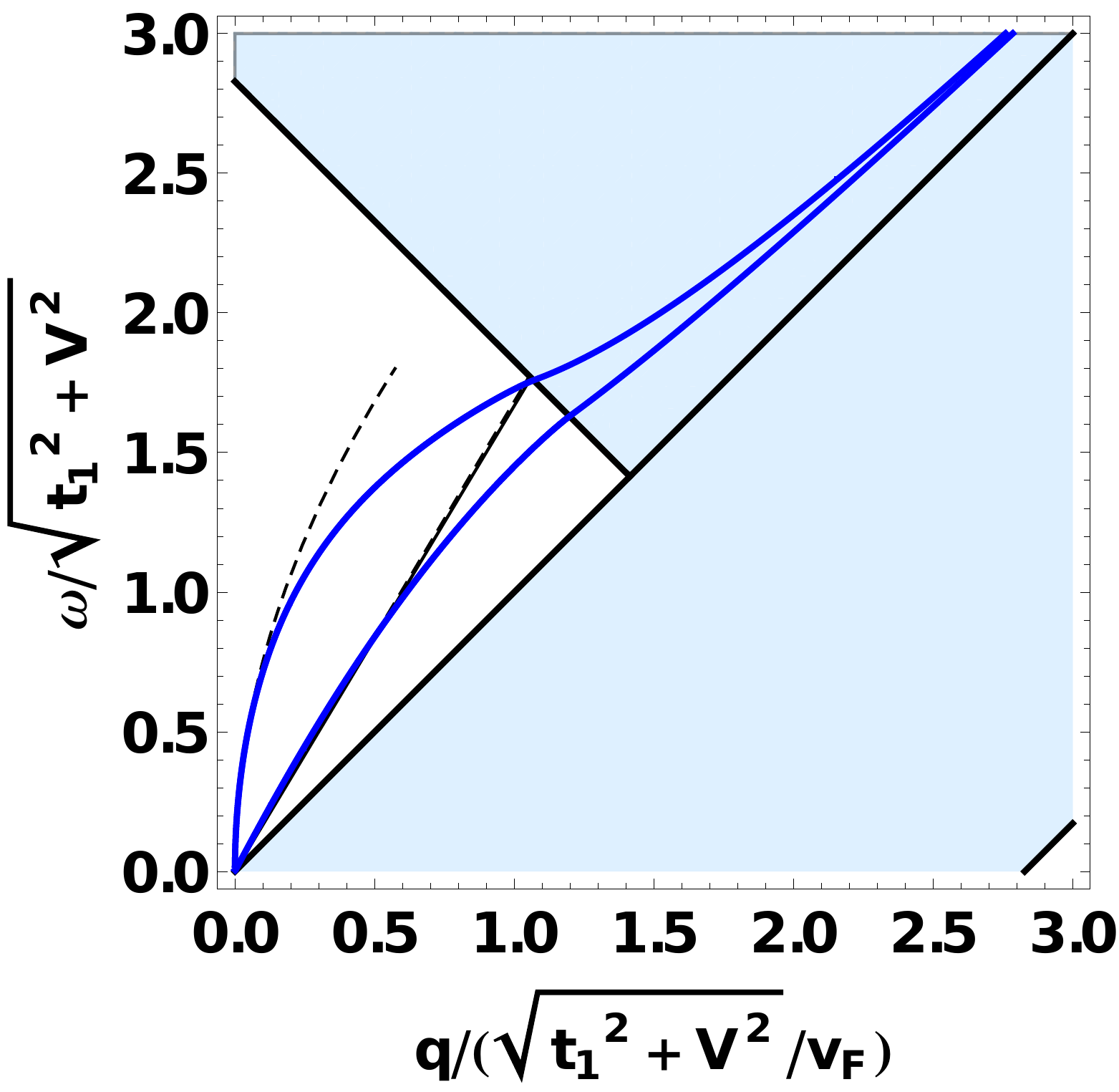}
\includegraphics[width=4.2cm]{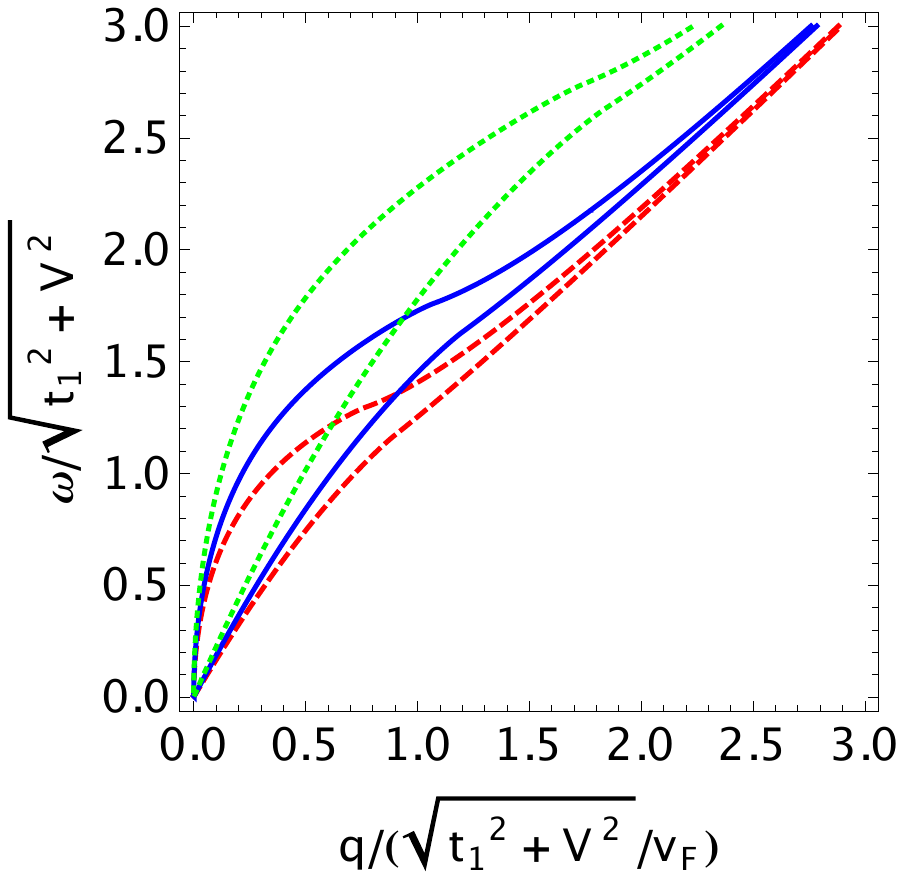}
} 
\end{center}
\caption{(Color online) Left: Numerical solution for he dispersion relation of the plasmon modes of AA-BLG in the presence of an external potential $V=t_1$ (solid blue lines) and their fitting to the analytical low energy expressions (\ref{Eq:PlasmonV1}) and (\ref{Eq:PlasmonV2}) (dashed black lines). The approximated dispersion relation of the acoustic mode given by (\ref{Eq:PlasmonAcVFull}) is shown by the full black line. Right: Dispersion relation of the plasmon for three different values of $V$: $V=2t_1$ (dotted green lines), $V=t_1$ (solid blue lines) and $V=t_1/4$ (dashed red lines).}
\label{Fig:PlasmonsV}
\end{figure}

The presence gate voltage in the AA-BLG creates a potential $+V$ in the top layer and $-V$ in the bottom layer. In this situation we can write the single-particle Hamiltonian as
\be
 H_V(\bk)
=
\left( 
\begin{array}{cc} 
{\cal H}^{+}_{{\rm SL}}(\bk) & {\cal H}_{\perp} \\
 {\cal H}_{\perp}  & {\cal H}^{-}_{{\rm SL}}(\bk) 
\end{array} 
\right)
\label{Eq:HAA-V}
\ee
where in this case
\be
{\cal H}^{\pm}_{{\rm SL}}(\bk)
=
\vf\left( 
\begin{array}{cc} 
\pm V & k_{-}  \\
k_{+} & \pm V 
\end{array} 
\right)
\ee
and ${\cal H}_{\perp}$ was given in Eq. (\ref{Eq:Hperp}). By applying to $H_V$ the same transformation as in Sec. \ref{Sec:Hamiltonian}, we can write the Hamiltonian (\ref{Eq:HAA-V}) in the form
\be
{\cal H}(\bk)
=
\left( 
\begin{array}{cccc} 
-t_1 & v_Fk_{-} & -V & 0 \\
 \vf k_{+} & -t_1 & 0 & -V \\
 -V & 0 & t_1 & \vf k_{-} \\
 0 & -V & \vf k_{+} & t_1
\end{array} 
\right),
\label{Eq:HAA2-V}
\ee
which admits the eigenenergies
\be\label{Eq:EnergyAA-V}
\varepsilon^V_{\lambda,s}(\bk)=s\sqrt{t_1^2+V^2}+\lambda v_F k
\ee
where $s=\pm1$ and $\lambda=\pm1$, and the eigenfunctions 
\be
\Psi^V_{s,\lambda}(\bk)
=
\frac{1}{\sqrt{1+\beta^2_s(t_1,V)}}
\left( 
\begin{array}{c} 
\beta_s(t_1,V) \psi_{\lambda}(\bk) \\ 
\psi_{\lambda}(\bk) 
\end{array} 
\right)
\ee
where $\psi_{\lambda}(\bk)$ has been given in Eq. (\ref{Eq:psi_lambda}) and 
\be
\beta_s(t_1,V)=\frac{t_1}{V}+s\sqrt{1+\frac{t_1^2}{V^2}}.
\ee

Also for this case, it is important to notice that the overlap of the electron and hole wave-functions ${\cal F}^V_{ss';\lambda\lambda'}(\bk,\bkp)=|\langle\Psi^V_{s,\lambda}(\bk)|e^{i(\bk-\bkp)\cdot \br}|\Psi^V_{s',\lambda'}(\bkp)\rangle|^2$ is
\be\label{Eq:FV}
{\cal F}^V_{ss';\lambda\lambda'}(\bk,\bkp)
 = \left\{ \begin{array}{ccc}
\frac{1+\lambda\lambda'\beta^2_s(t_1,V)\cos(\phi_{\bk}-\phi_{\bkp})}{2[1+\beta^2_s(t_1,V)]}& \mbox{for} & s=s' \\ 
  0  & \mbox{for} & s\ne s' 
\end{array}\right.
 \ee
 Therefore, because from Eq. (\ref{Eq:FV}), ${\cal F}^V_{ss';\lambda\lambda'}(\bk,\bkp)=0$ for $s\ne s'$, then only electron-hole transitions between bands with the same index $s$ contribute to the polarizability, as in the $V=0$ case.
 
Following the same scheme as in Sec. \ref{Sec:CE}, we find the dispersion relation for plasmons. The low energy dispersion relation for the optical mode in this case changes to
\be\label{Eq:PlasmonV1}
\omega_{+}(q)\approx \left(\frac{e^2g_{\sigma}g_v\sqrt{t_1^2+V^2}}{\kappa}q\right)^{1/2},
\ee
whereas the acoustic plasmon is
\be\label{Eq:PlasmonV2}
\omega_{-}(q)\approx \left(\frac{e^2g_{\sigma}g_vd\sqrt{t_1^2+V^2}}{2\kappa}\right)^{1/2}q.
\ee

In the left panel of Fig. \ref{Fig:PlasmonsV} we show the numerical solution for the plasmons in the RPA (solid blue lines) together with the long wavelength approximations (\ref{Eq:PlasmonV1}) and (\ref{Eq:PlasmonV2}), as shown by the dashed black lines. Notice that the velocity of the mode (the slope of the corresponding dispersion relation) can be controlled by tuning the strength of the external potential $V$. This effect is shown in the right panel of Fig. \ref{Fig:PlasmonsV}, where the dispersion of the modes for different strengths of the external field are shown. Those results are in agreement with the numerical calculations of AA-stacked multilayer graphene of Ref. \onlinecite{LW12}.

Finally, as we have discussed in Sec. \ref{Sec:Plasmons}, we notice here that the dispersion relation (\ref{Eq:PlasmonV2}) is only valid for
\be\label{Eq:Condition}
\frac{\sqrt{t_1^2+V^2}\,d}{v_F\kappa} \gg 1.
\ee 
A more accurate calculation of the long wavelength limit of the acoustic mode, following the method of Refs. \onlinecite{SG88} and \onlinecite{PM12}, gives a more accurate expression for the dispersion relation
\be\label{Eq:PlasmonAcVFull}
\omega_{-}(q)=\frac{1+\frac{g_{\sigma}g_v\sqrt{2}e^2d\sqrt{t_1^2+V^2}}{2\kappa v_F^2}}{\left(1+\frac{g_{\sigma}g_v\sqrt{2}e^2d\sqrt{t_1^2+V^2}}{\kappa v_F^2}\right)^{1/2}}\, v_Fq.
\ee
The fitting to the numerical dispersion relation obtained from Eq. (\ref{Eq:PlasmonAcVFull}) is shown by the full black line in the left panel of Fig. \ref{Fig:PlasmonsV}. Here it is interesting to notice that both long wavelength approximations, (\ref{Eq:PlasmonV2}) and (\ref{Eq:PlasmonAcVFull}), almost coincide for this case. This is due to the fact that the presence of a finite bias $V$, which acts as an effective doping of the sample, helps the condition (\ref{Eq:Condition}) for the applicability of the approximation Eq. (\ref{Eq:PlasmonV2}) to be fulfilled.

\subsection{Drude weight and plasmons in doped AA-BLG}\label{Sec:Doped}

In this section we study the characteristics of the plasmon dispersion of AA-BLG in the presence of doping. For this aim, we exploit the fact that the Drude weight of the whole system controls the dispersion relation of the {\it optical}~  $\omega_+(q)\sim q^{1/2}$ mode at long wavelengths.\cite{AM11} Indeed, in the $v_Fq\ll \omega \ll 2\mu$ limit ($\mu$ being the chemical potential), the real part of the interacting polarization function $\chi(q,\omega)=\Pi^0(q,\omega)/\epsilon(q,\omega)$  can be approximated by
\be\label{Eq:chi}
{\rm Re}\, \chi(q,\omega)\approx \frac{{\cal D}q^2}{\pi e^2 \omega^2}
\ee
where $\cal D$ is the Drude weight. Furthermore, $\cal D$ can be obtained from the optical conductivity $\sigma(\omega)$, which in the dynamical limit is simply ${\rm Im}\, \sigma(\omega) \approx {\cal D}/\pi \omega$. As a result, Eq. (\ref{Eq:chi}) leads to a solution for the plasmon dispersion of SLG of the form $\omega(q)=\sqrt{2{\cal D}q/\kappa}$. Furthermore, in Ref. \onlinecite{AM11} it was shown that, if vertex and self-energy corrections are neglected, the RPA Drude weight of a SLG is ${\cal D}=g_{\sigma}g_v\mu \sigma_0$, where $\sigma_0=e^2/4\hbar$ is the usual background conductivity. 

In a doped AA-stacked graphene bilayer, as a consequence of the characteristic electronic dispersion (\ref{Eq:EnergyAA}) and DOS (\ref{Eq:DOS}), the Drude weight has two independent contributions, each of them associated to one cone in the band structure:\cite{TN12}
\bea
{\cal D}={\cal D}_1+{\cal D}_2&=&g_{\sigma}g_v|t_1-\mu|\sigma_0+g_{\sigma}g_v(t_1+\mu)\sigma_0\nonumber\\
&=&2g_{\sigma}g_v\max(t_1,\mu)\sigma_0.
\eea

Therefore, the dispersion relation of the optical plasmon in a doped bilayer is, in the long wavelength limit
\be
\omega_+(q)\approx \sqrt{\frac{g_{\sigma}g_ve^2\max(t_1,\mu)}{\kappa}q}.
\ee
It is important to notice that, for values of doping such that $|\mu| \le t_1 \sim 0.2$~eV, which corresponds to carrier densities of the order of $n\sim 1.2\times 10^{13}$~cm$^{-2}$, the plasmon dispersion is completely independent of doping, $\omega_+(q)=\sqrt{(g_{\sigma}g_ve^2t_1/\kappa)q}$. This makes the plasmons in AA-BLG of special interest because their dispersion relation is {\it protected} against unintentional doping, due e. g. to charged impurities in the substrate. 

\section{Discussion}\label{Sec:Discussion}

In this section we summarize the main characteristics of the collective modes in AA-BLG, and discuss their differences with respect to the plasmons in other 2D systems. We consider first the most stable AB-stacking, which presents a completely different band structure as compared to AA-BLG, with energy dispersion $\varepsilon_{s,s'}(q)=s\sqrt{v_F^2q^2+t_{\perp}^2/4}+s' t_{\perp}/2$, where $s,s'=\pm 1$ and $t_{\perp}$ is the inter-layer hopping between A and B atoms of different layers.\cite{MK13} It contains four parabolic bands with effective mass $m=t_{\perp}/2v_F^2$: two of them that touch each other at the zero energy Dirac point, plus other two with maxima and minima at $\mp t_{\perp}$, respectively.\cite{CC10} The absence of carriers in the undoped case leads to an electron-hole continuum in which only inter-band excitations are allowed, preventing the existence of long-lived plasmons. This is also the case of SLG, whose band structure consists of two Dirac cones which touch at $\varepsilon=0$. In those two cases, AB-BLG and SLG, there is no phase space available for the propagation of coherent plasmons above the threshold of inter-band transitions. Therefore, collective modes are allowed to exist only in the doped regime.\cite{S86,WSSG06,HS07,P09,RFG13,G11} Completely different is the case of AA-BLG discussed here, for which long-lived plasmons are present even in the absence of any doping. Furthermore, the dispersion relation of the optical mode of AA-BLG, $\omega_+(q)=\sqrt{\frac{e^2g_{\sigma}g_vt_1 }{\kappa}q}$, is not affected by doping, unless the chemical potential exceeds the inter-layer hopping energy $|\mu|\ge t_1$. For higher doping, the optical plasmon recovers the well know dispersion relation of AB-BLG,  $\omega(q)=\sqrt{\frac{e^2g_{\sigma}g_v\mu }{\kappa}q}$.\cite{BM09,SHS10,G11} 

On the other hand, AB-BLG presents additional gapped high energy modes, which are highly damped since they lie entirely in the inter-band zone of the electron-hole continuum. The first of those modes, which has its origin in inter-band particle-hole transitions from the low energy band (with energy $\varepsilon(0)=0$) to the high energy band (with energy $\varepsilon(0)=t_{\perp}$), was studied by Gamayun in Ref. \onlinecite{G11}, who found the dispersion relation $\omega(q)\sim t_{\perp}+\frac{e^2g_{\sigma}g_v}{2\kappa}q\ln(1+2\mu/t_{\perp})$. A higher energy mode, with a gap of $\Delta=2t_{\perp}$, is also expected, associated to inter-band transitions between the low energy band with energy $\varepsilon(k)<-t_{\perp}$, and the high energy band with $\varepsilon(k)>+t_{\perp}$.\cite{NC08} Those gapped and damped modes, which are allowed in AB-BLG, does not have their counterpart in AA-BLG. The reason is that the completely decoupled bonding and antibonding sectors in AA-BLG forbids electron-hole excitations of the same nature, and therefore no gapped modes are expected to exist in AA-BLG. 

It is also interesting to compare our results to those of a coupled 2DEG-bilayer structure.\cite{SH98} Here the differences are quite significant as well. In fact, whereas in both cases there is a coexistence of a $\omega_+\sim q^{1/2}$ mode and a $\omega_-\sim q$ mode, the {\it acoustic} linearly dispersing mode becomes gapped in the 2DEG-BL when the effect of inter-layer tunneling is included.\cite{SH98} However, we have seen that in  AA-BLG the acoustic mode remains ungapped even in the presence of inter-layer tunneling. It is interesting to notice that the gap opened in the 2DEG-BLG when inter-layer hopping $t_{\perp}$ is considered, with a size of $\Delta=2t_{\perp}$, is the counterpart of the above mentioned gapped mode in AB-BLG. Therefore, whereas inter-layer tunneling opens a gap in the dispersion relation of the acoustic mode in 2DEG-BLG, the acoustic mode remains ungapped in AA-BLG whereas AB-BLG presents both, gapped and ungapped linearly dispersing collective modes.  

\section{Conclusions}\label{Sec:Conclusions}

In conclusion, we have studied the screening properties of AA-BLG. For this, we have calculated the dynamical polarization function, including the electron-electron interactions in the RPA. We have shown that the static screening in AA-BLG is similar to that of SLG, containing a metallic contribution (which is dominant at long wavelengths) and an insulating one (which dominates at large wave-vectors). However, whereas the metallic screening in SLG depends on the chemical potential (doping), here we have shown that in AA-BLG this metallic contribution is controlled uniquely by the inter-layer hopping energy. 

We have also studied the collective excitations of AA-BLG. We have obtained analytic low energy dispersion relations for the acoustic $\omega_-(q)\sim q$ and for the optical $\omega_+(q)\sim q^{1/2}$ plasmon modes. Our main result is that, due to the characteristic band structure and electron-hole continuum of AA-BLG, long-lived plasmons can disperse in AA-BLG {\it even} in the absence of doping. This is significantly different to SLG and AB-BLG, for which it is necessary to induce charge carriers in the sample to have coherent plasmon excitations. Furthermore, we have shown that the dispersion relation of the optical plasmon in AA-BLG does not depend on the amount of doping, unless high carrier concentrations with $|\mu| > t_1$ are reached. The effect of an electric field perpendicular to the sample has been considered, and we have shown that the dispersion of those modes can be manipulated by the application of a gate voltage.

\section{Acknowledgements}
The authors thank T. Stauber and P. San Jos\'e for helpful discussions. R. R. acknowledges financial support from the Juan de la Cierva Program and from grant FIS2011- 23713 (MINECO, Spain). L. B. acknowledges financial support by MINECO-Spain under grant FIS2012-33521.

\bibliography{BibliogrGrafeno}

\end{document}